\documentstyle[pre,aps]{revtex}
\begin{document}
\title{Orientational transitions in a nematic confined by competing surfaces}
\author{I. Rodr\'{\i}guez-Ponce \cite{corresponding}, J. M.  Romero-Enrique 
and L. F. Rull
\\
Departamento de F\'{\i}sica At\'omica, Molecular y Nuclear,
Area de F\'{\i}sica Te\'orica, Universidad de Sevilla, Aptdo 1065, 
E-41080 Sevilla, Spain}
\maketitle
\begin{abstract}
The effect of confinement on the orientational structure of a  
nematic liquid crystal model has been investigated by using a version of
density-functional theory. We have focused on the case of a nematic confined
by opposing flat surfaces, in slab geometry (slit pore), which favor
planar molecular alignment (parallel to the surface) and homeotropic
alignment (perpendicular to the surface), respectively.
The spatial dependence of the tilt angle of the director with respect to the 
surface normal has been studied, as well as the tensorial order parameter 
describing the molecular order around the director. For a pore of given width,
we find that, for weak surface fields, the alignment of the nematic director
is perpendicular to the surface in a region next to the surface favoring 
homeotropic alignment, and parallel along the rest of the pore, with a sharp 
interface separating these regions (S phase). For strong surface fields, 
the director is distorted uniformly, the tilt angle exhibiting a linear 
dependence with the distance normal to the surface (L phase). Our calculations 
reveal the existence of a first-order transition between the two director 
configurations, which is driven by changes in the surface field strength, and 
also by changes in the pore width. In the latter case the transition occurs, 
for a given surface field, between the S phase for narrow pores and the L 
phase for wider pores. A link between the L-S transition and the anchoring
transition observed for the semi-infinite case is proposed. 
\end{abstract}
\section{Introduction}
In the last years there has been a vast theoretical effort to 
understand the properties of liquid crystals in bulk phases. However, it is 
only recently that the interfacial phenomena presented by these complex 
fluids have attracted an increasing interest. The ability to manipulate the 
direction of the preferred molecular alignment, the {\it director}, by 
coupling to surfaces, is a property with a great practical interest. For 
this reason, there is a new effort trying to describe this rich {\it 
surface phenomenology} in the liquid crystals. These studies have mainly used
phenomenological Landau-de Gennes approaches and aimed to 
describe, with more or less success, different phenomena such as subsurface 
deformations \cite{Saupe,Barbero,Zumer} (ordering close to a surface), 
changes in the alingnment of the director due to the interaction with a solid 
substrate \cite{Poniewierski}, nematic-isotropic transition \cite{Robledo}, 
wetting \cite{Sullivan2}, etc. On the other hand, microscopic-type theories 
claim to provide a more precise description of theses inhomogeneous problems 
since they take account of the structure at the molecular scale. In this 
framework, there is a recent literature describing attemps to answer such 
questions and it is supposed to provide more satisfactory theories in the  
interfacial phenomena. For example, by employing an Onsager-type theory R. 
van Roij {\it et al.} \cite{Evans} investigated the problem of biaxiality 
and wetting for the Zwanzig model. P. I. C Teixeira has discussed, in a recent 
work \cite{Teixeira}, the existence of subsurface deformations for the case
of a confined nematic phase by using a generalized van der Waals theory. 
In previous papers, also employing this last approach, we studied 
surface phenomena in the case of a nematic in contact with a single wall 
\cite{Inma1} as well as capillary effects in the
case of a nematic in slab geometry with symmetric walls \cite{Inma2}.
In Ref. \cite{Inma1} our attention was restricted to the study of a nematic in
the presence of a solid interface. Model parameters where chosen such that the 
surface forced the molecules to lie perpendicular to the interface (homeotropic
orientation) while in the nematic-isotropic interface the nematic director 
had a preference to lie parallel to the interface (planar orientation). 
In the model, surface parameters responsible for the order are the 
particle-wall interaction ($\epsilon_W$) and the particle-particle interaction
($\epsilon_C$) which played an important role in the 
interplay between anchoring and wetting transitions. A first-order
anchoring transition between planar and homeotropic regimes was found;
in a $\mu-T$ (chemical potential-temperature) representation (see Fig. 1) 
this transition appears as a line called anchoring line. 
This line approaches the nematic-isotropic transition line tangentially 
(at constant $\mu$, eventually cuts this line at $T_D$), due to the existence 
of total wetting state at the coexistence. Later, the nematic phase was 
confined by symmetric walls, i.e both walls favored the same orientation 
(homeotropic) \cite {Inma2}. Then, also the same anchoring line appeared and 
it was found to be very insensitive to the pore width except with respect to 
the localization of $T_D$ that is shifted by capillary effects. 
These findings have a bearing on the new phenomenology which appears
when the liquid crystal is confined.

In the present paper, the theory is generalized to the
case where the walls compete in molecular alignment. We
believe that deeper insight into the surface-induced effects in liquid 
crystals can be obtained from a molecular approach and, in this respect,
density funtional theory appears to be a powerful tool. 
The confinement of the nematic phase is imposed by the 
presence of a boundary made up of two opposing flat surfaces,
one favoring homeotropic orientation and the other favoring planar orientation. 
In these conditions, a spatial variation in the tilt angle between the
director and the surface normal is found. The textures adopted by
the nematic director depend very dramatically on the values of both 
surface field and pore width. For the case of a
strong surface field $\epsilon_W$ and a wide pore this dependence is 
linear with respect to the coordinate normal to the interfaces
except close to the surfaces.
This configuration satisfies surface boundary conditions but entails
an elastic energy which is minimum for a linear tilt configuration.
For weaker surface fields and/or narrow pores a different texture 
occurs where the tilt angle adopts a steplike configuration. 
Our calculations also reveal the existence of
biaxial behavior close to the surface favoring planar orientation as 
well as in a small region associated to the change of orientation in 
the steplike configuration. 
These biaxilities are direct surface effects due to the existence of
different interfaces. 

In the next section we give a short account of the theoretical
model used to describe the structure and thermodynamics of the confined
liquid crystal. In section \ref{Results} we 
show the different tilt configurations
that result and argue that the phase transition between the linear-tilt 
and the steplike-tilt configurations can be related with the anchoring 
transition found in the semi-infinite problem. We also discuss how 
biaxiality is affected by the pore width. We end up with a section of
conclusions.

\section{The Model}
The theoretical model is a standard generalized
van der Waals theory based on a perturbative expansion, 
using a hard-sphere (HS) fluid
as reference system \cite{Margarida}.
Details on the physical basis of the model and
how to obtain its solutions numerically can be found elsewhere
\cite{anchoring,minimo}.
Our starting point is the grand potential functional
per unit system area $A$, $\Omega[\rho]/A$,
whose functional minimum with respect to the one-particle
distribution function, $\rho({\bf r},\hat{\bf\Omega})$, which depends
on both molecular positions ${\bf r}$ and orientations $\hat{\bf\Omega}$,
gives the equilibrium structure of the interface.
This function, $\rho({\bf
r},\hat{\bf\Omega})\equiv\rho(z)f(z,\hat{\bf\Omega})$, contains a mass
distribution $\rho(z)$ and an angular
distribution $f(z,\hat{\bf\Omega})$.
These quantities vary locally with the distance from $z=0$ to $z=H$, $H$
being the pore width.
The expression for $\Omega[\rho]$, in a mean field approximation is,
\begin{eqnarray}
&&\Omega[\rho]=F_{r}[\rho]+\frac{1}{2}\int\!\!\int\!\!\int\!\!\int d{\bf r}
d{\bf r}^{\prime}d\hat{\bf\Omega}d\hat{\bf\Omega}^{\prime}
\rho({\bf r},\hat{\bf\Omega})\rho({\bf r}^{\prime},\hat{\bf\Omega}^{\prime})
\nonumber\\
&&\hspace{0cm}\times v({\bf r}-{\bf r}^{\prime},\hat{\bf\Omega},
\hat{\bf\Omega}^{\prime})-\int\!\!\int d{\bf r}d\hat{\bf\Omega}
\rho({\bf r},\hat{\bf\Omega})[\mu-v_W({\bf r},\hat{\bf\Omega})],
\end{eqnarray}
where $\mu$ is the chemical potential and
\begin{eqnarray}
F_{r}[\rho]=\int d{\bf r}f_{hs}(\rho({\bf r})) + 
k_BT\int d{\bf r}\rho({\bf r})<\ln(4\pi f(z,\hat{\bf \Omega}))>             
\nonumber
\end{eqnarray}
is the reference system free energy. In the above expression
${f}_{hs}(\rho({\bf r}))$ is the hard-sphere free energy
density of a uniform fluid with a density equal to the local density at 
${\bf r}$ and '$<...>$' is an angular average. The attractive potential
$v$ contains anisotropic (dispersion) forces driving the
liquid-crystalline behavior of the model material:
\begin{eqnarray}
v({\bf r},\hat{\bf\Omega},\hat{\bf\Omega}^{\prime})=
v_A(r)+v_B(r)P_2(\hat{\bf\Omega}\cdot\hat{\bf\Omega}^{\prime})
+v_C(r)\left[P_2(\hat{\bf\Omega}\cdot{\hat{\bf r}})+
P_2(\hat{\bf\Omega}^{\prime}\cdot{\hat{\bf r}})\right],
\label{pot}
\end{eqnarray}
where $\hat{\bf r}={\bf r}/r$, and $v_A(r)$, $v_B(r)$, $v_C(r)$ are
functions of the
intermolecular center-of-mass distance $r$. Note that this potential 
is adequate to model uniaxial molecules with top-bottom symmetry. In 
this work we choose them
to have a simple Yukawa form, i.e. \
$v_i(r)=-\epsilon_i\exp{(-\lambda_i (r-\sigma))}/r$ for
$r>\sigma$, and $v_i(r)=0$ otherwise, where $\sigma$ is the diameter
of a hard sphere.

The walls are modeled via the following potentials:
\begin{equation}
v^1_W(z,\theta)=-\epsilon_W
e^{-\lambda_W (z-\sigma)}P_2(\cos{\theta})\hspace{0.8cm} 
\hbox{(left wall,}\hspace{0.1cm}z=0)
\end{equation}                                            
which favors homeotropic anchoring, and 
\begin{equation}
v^2_W(z,\theta)=\epsilon_We^{-\lambda_W (-(z-H)-\sigma)}
P_2(\cos{\theta})\hspace{0.8cm}
\hbox{(right wall,}\hspace{0.1cm}z=H)
\end{equation}
which favors planar alignment. The parameter $\epsilon_W$ is the 
surface strength of the walls and plays an important role in 
anchoring phenomena. 
  
In order to describe the orientational structure of the fluid a 
tensorial order parameter is defined as,  
\begin{eqnarray}
Q_{\alpha \beta}=\int d\hat {\bf \Omega} f(z,\hat {\bf \Omega})
\left( {3\hat \Omega_\alpha \hat \Omega_\beta - \delta_{\alpha \beta}
\over 2}\right)
\end{eqnarray}
where $\delta_{\alpha \beta}$ is the Kronecker symbol and
$\hat \Omega_\alpha$ is the $\alpha$-component of $\hat {\bf \Omega}$. 
This quantity is a traceless symmetric tensor with eigenvalues $\lambda_1 \ge
\lambda_2 \ge \lambda_3$. In this description, the isotropic state is 
defined by the condition $\lambda_1=\lambda_2=\lambda_3=0$. Uniaxial 
states occur when there is a twofold degenerated eigenvalue and then 
two cases can be distinguished: uniaxial nematic state ($\lambda_1>\lambda_2=
\lambda_3$) and random planar ($\lambda_1=\lambda_2>\lambda_3$). In the 
first case, the molecules align preferently along the director, defined as 
the eigenvector corresponding to the largest eigenvalue (the
uniaxial order parameter $U\equiv \lambda_1$).  
In the second case, the molecular orientations are uniformly distributed on 
a plane perpendicular to the eigenvector correponding to the lowest 
eigenvalue.  

For the case of non-degenerated eigenvalues, we can define the 
uniaxial ($U$) and biaxial ($B$) order parameters in terms of the 
eigenvalues of this tensor. The former is defined, as before, as the largest 
eigenvalue of the order parameter tensor ($\lambda_1$) and the 
corresponding eigenvector is the local director of the fluid. 
The biaxial parameter is proportional to the difference of the two smallest 
eigenvalues. This quantity gives information of the amount of the 
orientational order on the plane 
perpendicular to the director. In this way, a 
secondary director can be defined as the eigenvector 
corresponding to the eigenvalue $\lambda_2$.  
By the symmetry of the problem, 
we have focused in the case where the director  
varies in the $xz$ plane and is invariant with respect to a reflexion 
$y\rightarrow-y$. 
In these conditions, the director is characterized by the tilt angle $\psi$, 
defined as the angle formed between the director and the $z$ axis. 
By convention, we define $B=\pm 2(\lambda_2-\lambda_3)/3$, with 
$B > 0$ if the secondary director is on the $xz$ plane and 
$B < 0$ if it is along the $y$ axis. 

For numerical reason, it is convenient to obtain          
the orientation distribution of the molecules referred to a laboratory 
reference system, described by three order parameters,
\begin{eqnarray}
\eta(z)=\int d\phi \sin \theta d\theta f(z,\hat \Omega)P_2(\cos \theta) \\
\nu(z)=\int d\phi\sin \theta d\theta f(z,\hat \Omega)\sin2 \theta\cos \phi \\
\sigma(z)=\int d\phi\sin \theta d\theta f(z, \hat \Omega)\sin^2 \theta \cos2 
\phi
\end{eqnarray}
These independent parameters can be related to the set $\{\psi, U, B\}$ by 
the following expressions, 
\begin{eqnarray}
&&\tan \psi(z)={\nu(z)\over{\eta(z)-\sigma(z)/2}+\sqrt{\left[\eta(z)-
\sigma(z)/2\right]^2+\nu(z)^2}}  \\
&&U(z)={1\over 4}\left ( \eta(z)+{3\over 2}\sigma(z)+3\sqrt{[\eta(z)-(1/2)\sigma(z)]^2+\nu(z)^2}\right )  \\
&&B(z)={1\over 2} 
\left (\eta(z)+{3\over2}\sigma(z)-\sqrt{[\eta(z)-
(1/2)\sigma(z)]^2+\nu(z)^2}\right).
\end{eqnarray}

Numerical values for the potential parameters were taken as
$\epsilon_A=1$ (which sets the temperature scale),
$\epsilon_B/\epsilon_A =0.847$ and $\epsilon_C/\epsilon_A=0.75$,
The range parameters $\lambda_i$ are set, in units of
$\sigma$ (throughout we choose this unit to set the length scale),
to $\lambda_i=2,4,1.75$, $i=A,B,C$ respectively, and $\lambda_W=1$.
The model predicts a bulk phase diagram with vapor, isotropic liquid
and nematic liquid coexisting at a triple point temperature $T_{NIV}$ 
\cite{Margarida}.

\section{Results}
\label{Results}
Confinement of simple liquids generally brings about capillary effects and
associated capillary condensation phenomena whereby condensation occurs
below the saturation point in bulk, as measured for example by the
chemical potential. Similar phenomena must occur in nematic
liquid crystals but with additional complicating factors due to the 
orientational order induced by the confining surfaces.

In the following we use the chemical potential $\mu$ as the external
thermodynamic potential, in addition to the temperature $T$. 
Capillary condensation of the nematic from the isotropic phase 
occurs when, at fixed $\mu$, the temperature $T_{NI}(H)$ at which
the pore becomes filled with an oriented (nematic) phase is shifted with
respect to the corresponding temperature in bulk $T_{NI}(\infty)$. 
We are thus seeking a non-zero value of $\Delta T\equiv
T_{NI}(H)-T_{NI}(\infty)$. Our calculations reveal that, contrary
to the case of symmetric walls \cite {Inma2} and, for large pore widths 
($H$ larger than $20\sigma$), no significant capillary effect 
exists, i.e. we find that the NI transition occurs at the same temperatures
as in the bulk state. However, there are indications that for smaller pore 
widths this situation is not true longer (this aspect is being studied 
currently). Note though that condensation lines 
involving confined vapor states are expected to be affected by 
confinement; since these lines are not central to our argument we do not
give any results in the following.

Fig. 1 shows the bulk phase diagram, 
with vapor (V), isotropic (I) and nematic (N) phases and their 
corresponding phase transitions. We have superimposed the anchoring
line, representing surface phase transitions between states with
different director alignment for the semi-infinite problem 
(i.e. only one surface, inducing
homeotropic alignment); these results have been obtained with a surface
field $\epsilon_W=0.53$, for larger values this transition curve tends to 
approach the coexistence lines reducing then the region of planar states 
\cite{Inma1}. 

Now, we have calculated equilibrium states but with the sample 
confined between two opposing walls separated a distance
$H$. 
There is a molecular ordering field due to the surfaces through the potential 
parameter $\epsilon_W$.
If the surface field strength is sufficiently high, one
expects the director to adopt the configuration which satisfies the
orientation favored by each surface, at least
approximately.
 However, this necessarily demands that the director
becomes distorted within the pore which entails a free energy cost due to 
 the elastic contribution. 
In fact, the distribution of this inhomogeneity along the pore
is found to depend sensitively on the values of pore width and
surface field strength, and is expected to also depend on the thermodynamic
parameters $\mu$ and $T$. 

Competition between these two energies results in the existence of two
tilt configurations: L, where the tilt angle rotates uniformly along
the pore and the resulting tilt-angle profile is linear in $z$;
S, where the tilt angle adopts a step-like configuration, being 
constant throughout the sample except in a narrow region close to the 
$z=0$ surface where a crossover exists from $0^\circ$ (homeotropic) to 
$90^\circ$ (planar).

Fig. 2 shows an L phase for $H=20\sigma$, $\epsilon_W=0.7$ and for 
thermodynamic conditions of temperature and chemical potential where, in the 
semi-infinite case, the nematic is clearly in the homeotropic state 
($\mu=-3.7$, $T=0.57$). In the middle of the pore the tilt profile is almost 
a linear function of $z$, approaching smoothly the walls to their favored 
values. 

In the Fig. 3 the density, the uniaxial and biaxial order parameter 
profiles are plotted. The profiles present a smooth behavior, reaching a 
clear plateau in the middle of the pore with values close to the bulk values. 
The vanishing value of $B$ in regions far from the interfaces is consistent 
with the uniaxial nature of the fluid particles. 
Close to walls the profile behaves differently depending on the 
favored orientation. Around the $z=0$ surface the $U$ order parameter 
increases due to the strong surface interaction (despite of a decay in 
density). However, these deformations do not induce biaxiality (i.e. $B=0$) 
due to the smooth character of the tilt 
profile along the pore.  
On the other hand, near the $z=H$ surface, where the tilt angle 
is $90^\circ$, the $U$ order 
parameter gets lower values and $B<0$, that reflects the orientation 
 of the molecules are mainly distributed on the $xy$ plane, being 
the $x$ axis the in-plane preferred direction. The decrease of 
the amount of order is due to 
surface potential that promotes a random planar distribution.   

If the surface field strength is
reduced the director tilt configuration changes over to a step, i.e. a S 
phase. Here the tilt angle is uniform and equal to $90^\circ$ except in a
narrow homeotropic region close to the homeotropic wall (Fig. 4). 
In general, this homeotropic region 
increases with the wall-particle potential range and with the strength of the 
surface field, so for the case of $\epsilon_W=0$ one recover the configuration 
which corresponds to a nematic-hard walls interfaces (planar throughout the 
pore). Note that the S phase is spatially asymmetric, the tilt orientation 
being different from planar only in a small region close to the 
homeotropic wall. In fact, a corresponding S phase where the tilt were
in a homeotropic configuration except in a very small region close to the
planar wall is not observed in our calculations because of the effect
due to the ordering field coming through the potential parameter $\epsilon_C$. 
This field promotes molecular configurations where the director lies
along a direction either perpendicular or parallel to the direction of 
inhomogeneity, depending on the sign of $\epsilon_C$ ($\epsilon_C>0$ or 
$\epsilon_C<0$, respectively). In our case $\epsilon_C>0$, which means that 
the preferred director orientation is planar except in the region, close to 
$z=0$, where the effect of the wall promoting homeotropic alignment is more
pronounced.

In the figure 5 we plot the density, uniaxial and biaxial order parameters 
profiles  in the S phase
obtained with $\epsilon_W=0.30$. Again the profiles only deviate from 
the bulk values close to the surfaces. 
The deviations around $z=H$ has the same origin that in the L phase. However, 
close to $z=0$ this behavior has different nature. 
An analysis of the order parameteres profiles reveals that the system 
goes {\it smoothly} from an homeotropic to a planar configuration via 
a random planar state in the $xz$ plane. This point ($z_1$ at fig. 5) 
corresponds to the jump of the tilt (origin of the step function profile). 
This mechanism is reponsible of the decay of the order and the biaxial 
behavior around this region. In fact, the existence of a crossover 
between homeotropic configuration and a planar configuration is closely 
linked to the capability of the fluid to have biaxial behavior in a small 
spatial range between both anchoring states. So, the roles that the $z$ and
$x$ axes play as the local director and secondary director, respectively, 
for $0<z<z_1$ are exchanged at $z=z_1$. However, in order to make this
mechanism feasible a dramatic depletion on the amount of order is needed.
Moreover, the secondary director changes its alignment from the $z$ axis to
the $y$ axis as $z$ is increased (i.e. $B$ changes its sign), before that 
biaxiality virtually disappears in the middle of the pore.
Biaxiality directly induced by the breaking of symmetry associated with 
the presence of the interfaces and, in the case of a free interface with 
planar director alignment, has been observed in simulations \cite{Cleaver}. 

In order to study in more detail the dependence of the nematic ordering on the 
pore width, we have calculated uniaxial order and biaxiality profiles for 
different values of $H$ and for $\epsilon_W=0.7$ (Fig. 6). For pore widths
$H=20, 30$ and $40$, for which the L phase is the stable phase, the profiles 
exhibit inhomogeneity regions close to the
surface favoring planar alignment, being almost constant in the rest of
the pore. Note that biaxiality occurs only where the tilt angle is
significantly close to $90^\circ$. An interesting feature is that the order
parameter profiles converge around to $z=0$ (resp. $z=H$) to the profiles
obtained for the $z=0$ surface against a nematic in a homeotropic 
configuration (resp. the $z=H$ surface against a nematic in a planar 
configuration). Furthermore, such semi-infinite profiles are the 
equilibrium ones for these values of $\epsilon_W$, $\mu$ and $T$.  
On the other hand, for $H=7.5$, the
stable phase is the S phase, and additional regions of inhomogeneity
(and biaxiality) develop in the neighbourhood of the $z=0$ surface. 

The transition between the L and S phases involves a finite free-energy
barrier, implying the existence of a thermodynamic first-order phase
transition between the two phases. This is revealed by computing the
relevant free energy for confined systems, i.e. the grand potential,
as a function of the potential parameters and/or the thermodynamic
parameters $\mu$ and $T$. In our case we have chosen to fix the latter
and vary $\epsilon_W$. Fig. 7 shows the behavior of the grand 
potential as a function of $\epsilon_W$. Two branches, corresponding
to the L and S phases, are shown. The point where the branches cross,
$\epsilon_W^c$, gives the transition point. Note the existence of 
metastable branches in both phases, which indicates the first-order nature of 
the phase transition. Tilt angle profiles along the pore corresponding 
to the coexisting L and S phases at $\epsilon_W=\epsilon_W^c=0.554$ for 
$H=20\sigma$ are shown in Fig. 8. Note that the linear behavior
in the L phase has an smaller range and its slope is also less than
in the case presented in Fig. 2. On the other hand, the step in the S 
phase is located further from the $z=0$ surface than in the case presented 
in Fig. 4.  

As in any analysis of metastable states, some caution
has to be exercised in analysing the occurrence of the different
phases. In particular, the initial conditions required in our numerical
minimisation scheme have to be chosen carefully. For example, the reality
of the metastable L branch is shown by performing the following
analysis: i) Equilibrium S states at low $\epsilon_W$ are obtained by
conducting minimisation processes starting from linear and step-like 
profiles; these processes give the same equilibrium S states. ii) 
For higher, increasing values of $\epsilon_W$, using step-like profiles or
linear profiles as starting conditions give different final states
beyond the transition point $\epsilon_W^c$, the final L states having lower
grand potential energies than the final S states. This is the region
of metastability of the S phase. Proceeding in the same way but from the L 
branch at high $\epsilon_W$ and decreasing $\epsilon_W$ provides the 
metastability region for the L phase.

Fig. 9 shows the global phase diagram with respect to the parameters 
$\epsilon_W$ and $H$ and for thermodynamic conditions $T=0.57$ and
$\mu=-3.7$. The line indicates the phase transition separating L and
S phases. As the pore width is increased, the region of stability of the 
L phase is enlarged since the elastic energy associated with the 
deformation of the director along the pore (proportional to the
integrated squared gradient of the tilt profile) decreases as opposed
to the case of the S phase, for which this energy is essentially constant
and seems to approach asymptotically to the value of 
$\epsilon_W=\epsilon_W^{a}$, at which the anchoring transition occurs 
for the semi-infinite case of the nematic against the $z=0$ wall.
For narrow pores, we find a quick increase of $\epsilon_W^c$ 
as the pore width is reduced, indicating that, in
order for the L phase to become stable, a large surface field is required to
compensate for the large elastic energy associated with a high tilt angle
gradient. An interesting issue, which we have been unable to address due to
technical reasons, is whether this first order phase transition may end up at
a critical point as the pore width is reduced or even become second order 
through the appearance of a tricritical point.

An analysis of the results quoted above shows that there is a relationship
between the L-S transition and the anchoring transition observed in
the semi-infinite problem. Let us consider the case of $H \gg \sigma$.
We have observed that the density and order parameter profiles deviate
from the bulk values only in regions around of the surfaces, being
the ranges of the inhomogeinities of the order of the molecular interactions. 
This feature is not surprising for the S phase, since in the middle of the
pore the nematic is in a non-distorted tilt configuration (i.e. planar),
so its tilt profile can be understood as two uncorrelated semi-infinite 
profiles linked via the bulk planar state. However, this fact is also 
true for the L phase even when the tilt distorsion is propagated through
the pore (see discussion for the Fig. 6 above). This is the scenario 
that the phenomenological theory implicitly assumes: the excess free energy 
$\Delta \Omega = \Omega + p V$ (where $p$ is the bulk pressure and $V = A H$ 
is the volumen of the sample) can be split in 
two terms, an elastic term $\Delta \Omega^e$ that comes from the 
distortions of the director field in a macroscopic scale, and surface terms,
$\Delta \Omega^{surf}_{z=0}$ and $\Delta \Omega^{surf}_{z=H}$, that 
include the effect of the density and order parameter inhomogeneities in
a microscopic scale close to the boundaries. The elastic term is given in
terms of the elastic constants $K_1$, $K_2$, $K_3$ and $K_{24}$ by
the Oseen-Frank free energy \cite{Oseen,Frank} as:
\begin{equation}
\Delta \Omega^e = {K_1 \over 2} ({\bf \nabla \cdot n})^2 + 
{K_2 \over 2} ( {\bf n \cdot \nabla} \times {\bf n})^2 +
{K_3 \over 2} ( {\bf n} \times {\bf \nabla} \times {\bf n})^2 -
(K_2 + K_{24}) {\bf \nabla \cdot}({\bf n \nabla \cdot n}+
{\bf n} \times {\bf \nabla} \times {\bf n})
\label{Oseen-Frank}
\end{equation}
where ${\bf n}({\bf r})$ is the director field. Note that the 
splay-bend term is not included since $K_{13}=0$ in bulk \cite{Yokoyama}. 
If we restrict our study to the cases in which there is translational
symmetry in the $xy$ plane and the director is assumed to be in the 
$xz$ plane (there is no twist), then ${\bf n}({\bf r}) = 
(\cos(\psi(z)),0,\sin(\psi(z))$. 
In our model of uniaxial molecules with top-bottom symmetry, 
$K_1=K_3=-2 \pi \rho_b^2 U_b^2 \int_0^\infty dr r^4 v_B(r)$, where 
$\rho_b$ and $U_b$ are the bulk density and
uniaxial order parameter, respectively \cite{TeixeiraJCP,Yokoyama}. 
So, the total excess free energy is given by:
\begin{equation}
{\Delta \Omega \over A} = {\Delta \Omega^{surf}_{z=0}(\psi(0),\psi'(0))
\over A} + {\Delta \Omega^{surf}_{z=H} (\psi(H),\psi'(H)) \over A}
+ {K_3 \over 2} \int_0^H dz (\psi'(z))^2
\label{phenomenologic}
\end{equation} 
where $\psi' \equiv d \psi / dz$. Note that the dependence of the surface
terms on the contact values at the boundaries of the tilt angle and
its derivative respect to $z$ \cite{Faetti}. Functional minimization
of (\ref{phenomenologic}) leads to an equilibrium tilt linear profile
$\psi' = constant = O(1/H)$. The exact expression for the  
profile will depend on the surface terms of the free energy, so it is 
needed some information about them. The values of $\Delta \Omega^{surf}_{z=0}$
can be obtained microscopically from the excess free energy corresponding
to a case of the wall against a distorted nematic whose tilt profile
goes asymptotically as $\psi(0) + \psi'(0) z$ for $\sigma \ll z \ll H$
(this discussion is analogous for the $z=H$ case). Since $\psi'(0)\ll 1/\sigma$ 
the effect of the distortion far from the $z=0$ surface will be a perturbation 
respect to the case $\psi'=0$, that corresponds to the semi-infinite case 
studied in \cite{Inma1}. In this Reference it was shown that if the wall
promotes homeotropic anchoring, depending on the ratio between $\epsilon_W$
and $\epsilon_C$ two local minima can appear for $\psi=0^\circ$ and
$\psi=90^\circ$ with a free energy barrier between them. However, 
if the surface favors a planar anchoring
only a global minimum at $\psi=90^\circ$ can exist. We can expand
now the surface free energy in a series of both $\psi$ and $\psi'$ around
$(\psi,\psi')=(0^\circ,0)$ and $(90^\circ,0)$. The special symmetry of 
our problem (rotational invariance around the $z$ axis and the
top-bottom symmetry of the molecules) implies that the linear term in
$\psi'$ vanishes. From this analysis we can conclude that for 
$H \gg \sigma$ two different states can exist: a linear state that goes
from a tilt close to $0^\circ$ at $z=0$ to a value close to $90^\circ$
at $z=H$, and a state with non-distorted planar tilt through the pore,
and that correspond to the L and S states we have found, respectively. 
Furthermore, by using this
phenomenologic approach a first-order transition can be predicted close
to the anchoring line, inside the homeotropic region. 
The equation (\ref{phenomenologic}), up to terms of the order of $O(1/H)$, is:
\begin{equation}
{\Delta \Omega \over A}= {\Delta \Omega^{surf}_{z=0} (\psi_0,0) \over A} 
+ {\Delta \Omega^{surf}_{z=H} (\pi/2,0) \over A} + {K_3 \over 2 H} 
\left({\pi \over 2}-\psi_0\right)^2
\label{approx}
\end{equation}
where $\psi_0=0$ or ${\pi/2}$ for the L and S states, respectively.
The transition occurs for
\begin{equation}
H_t= {\pi^2 K_3 \over 8 \left(\Delta \Omega^{surf}_{z=0}(\pi/2,0) - 
\Delta \Omega^{surf}_{z=0}(0,0)\right)}
\label{ht}
\end{equation}
which implies that $\Delta \Omega^{surf}_{z=0}(\pi/2,0) > \Delta 
\Omega^{surf}_{z=0}(0,0)$, i.e. the conditions should correspond to 
have an homeotropic configuration in the semi-infinite case, and
the transition is first-order due to the free-energy barrier that 
exists between the two minima for the semi-infinite case. So, 
for $H > H_t$ an stable L configuration exists and for $H<H_t$ the
S state is the stable one. From the Eq. (\ref{ht}) it is clear that
$H_t \to \infty$ as the system approaches to the anchoring line 
(characterized by the condition $\Delta \Omega^{surf}_{z=0}(\pi/2,0)=\Delta
\Omega^{surf}_{z=0}(0,0)$), i.e. the L-S transition 
occurs for infinitely wide pores and, consequently, the anchoring line 
can be understood as the limit of the L-S transition for $H \to \infty$.

Strictly speaking, the scenario presented above is only true for $H \gg 
\sigma$. However, our calculations reveal that the L-S transition
observed by our Density Functional Theory (DFT) calculations is the 
continuation of the transition predicted for wide pores. First of all, 
the free energy of the S phase is expected to converge towards its limiting 
value given by equation (\ref{approx}) very fast, as it was observed in the 
symmetric walls case in \cite{Inma2}. For the L phase, we have compared the 
DFT results for $\epsilon_W=0.7$ with the limiting expression given by 
the Eq. (\ref{approx})
(see Fig. 10). The values of $\mu$ and $T$ are the same as in all our
calculations, i.e. $\mu=-3.7$ and $T=0.57$. For such conditions, $K_3=0.615$.
On the other hand, calculations for the semi-infinite case show that
the asymptotic value for $H \to \infty$ is given by $\Delta \Omega / A K_B T=
1.841$. It is clear that the values of $\Delta \Omega$ converge for 
relatively small values of $H$ to the expression given by the phenomenologic
theory, and in any case the behavior is qualitatively correct. 
>From this analysis we conclude that the L-S transition is driven by
the anchoring (surface) transition that occurs for the semi-infinite case
of the nematic against the $z=0$ surface. 

\section{Conclusions}

In this paper we have employed a well-tested generalised van der Waals theory
to investigate the effect of confinement on the thermodynamics and 
the microscopic structure of a 
nematic liquid crystal. The presence of two confining surfaces favoring 
opposite orientations of the nematic director makes the order parameter,
which is a tensorial quantity, exhibit all of its richness, giving rise
to capillary effects not present in simple, non-orientable fluids. The 
competing conditions at the surfaces imply a deformed state of the director 
inside the pore.

The main result of our study is that the system may adopt two
possible states: one where the director changes orientation 
at an interface located close to the surface that favors homeotropic 
anchoring  (depending on the wall-particle potential range), 
and one where the deformation
takes place uniformly, giving rise to a basically linear tilt profile, with 
some degree of deformation close to the surfaces. The director may change from
one configuration to another via a first-order phase transition. The interplay 
between these two phases may be understood in terms of the competition between
the elastic energy associated with a deformed director state and the energy 
associated with the orienting surfaces. We have also shown that this  
transition is closely linked with the planar-homeotropic transition observed 
in the semi-infinite problem. In order to prove this aspect, a phenomenologic 
treatment has been employed. On the one hand, the theory shows how for 
big pores ($H \gg \sigma$) the L-S first order transition can occur at 
threshold value of the pore width $H_t$. 
On the other hand, the phenomenological approach also reveals that 
for $H\to \infty$ the L-S transition corresponds to the anchoring transition 
of the semi-infinite problem. We also conclude that the transition obtained
by the DFT is the continuation to shorter scales of the transition 
predicted by the phenomenological theory, that in principle is only
valid for much larger pore widths.  

\section*{Acknowledgments}
I.R.P. would like to thank Prof. R. Netz for his hospitality at LMU,
which helped to develop this work. This research was supported by grants 
No. PB97-0712 from DGICyT of Spain and No. FQM-205 from PAI of the Junta de 
Andaluc\'{\i}a.

\newpage

\begin{figure}
\caption{Phase diagram containing bulk phase transitions and the 
anchoring transition for the case $\epsilon_W=0.53$ (they join at the 
temperature $T_D$). NI coexistence is denoted by the thick solid line.
The dotted line is the anchoring line for the case of a 
nematic in the presence of a single surface which favors 
homeotropic orientation; this line
divides the nematic phase into a region with homeotropic ($\perp$) and 
planar ($\parallel$) nematic states. The thin solid line corresponds to the 
bulk transitions involving the vapor phase.}
\label{fig1}
\end{figure}                                                               

\begin{figure}
\caption{Tilt-angle profile for a confined nematic at $T=0.57$ and $\mu=-3.7$
(L phase). Values for the surface field and pore width are $\epsilon_W=0.7$ 
and $H=20\sigma$, respectively. 
}
\label{fig2}
\end{figure}
\begin{figure}   
\caption{Density ($\rho$), uniaxial ($U$) and biaxial ($B$) 
order parameter profiles 
for the L phase. $\epsilon_W=0.7$ and values of $T$ and $\mu$ are given 
in text.}
\label{fig3}
\end{figure}

\begin{figure}
\caption{Tilt-angle profile for a confined nematic with surface field 
$\epsilon_W=0.30$ (S phase). The values of $T$, $\mu$ and $H$ are the same 
as in Fig. 2.} 
\label{fig4}
\end{figure}

\begin{figure}   
\caption{Density($\rho$), uniaxial ($U$) and biaxial ($B$) order parameters 
profiles for the S phase. 
$\epsilon_W=0.3$ and the values of $T$ and $\mu$ are the same
as in Fig. 2.}
\label{fig5}
\end{figure}
\begin{figure}   
\caption{Uniaxial ($U$) and biaxial ($B$) order parameter profiles
for different pore widths
and $\epsilon_W=0.7$. The pore widths $H$ are, from left to right:
$7.5$, $20$, $30$, and $40$. The values of $T$ and $\mu$ are the same
as in Fig. 2.}
\label{fig6}
\end{figure}

\begin{figure}
\caption{Thermodynamic grand potential as a function of the surface 
field $\epsilon_W$ for thermodynamic conditions and pore width as in Fig. 2.
Branches corresponding to step-like (S) and linear (L) tilt-angle profiles
are shown. The point at which these branches cross gives the transition
point at $\epsilon_W^c=0.554$. The values of $T$, $\mu$, and $H$ are the same
as in Fig. 2.}
\label{fig7}
\end{figure}

\begin{figure}   
\caption{Tilt-angle profiles for L and S phases at coexistence 
corresponding to pore width $H=20$ and critical value $\epsilon_W^c=0.554$. The values of $T$ and $\mu$ are the same
as in Fig. 2.}
\label{fig8}                           
\end{figure} 

\begin{figure}   
\caption{Phase diagram in the $\epsilon_W-H$ plane.
The values of $T$ and $\mu$ are the same as in Fig. 2.}
\label{fig9}                           
\end{figure} 
\begin{figure}
\caption{Comparison between the Density Functional Theory and the
phenomenologic theory (expression in the text). The diamonds correspond 
to the excess free energy per area in units of $K_B T$ of a nematic confined 
by opposing walls in the L state with $T=0.57$, $\mu=-3.7$ and 
$\epsilon_W=0.7$, calculated by using the Density Functional approach. 
The solid line 
is the surface tension given by the phenomenologic theory  
given by the Eq. (\ref{approx}).}
\label{fig10}
\end{figure}


\begin{thebibliography}\

\begingroup
\bibitem[*] {corresponding} Corresponding author: inma@likix2.us.es
\bibitem{Saupe}
J. Nehring and A. Saupe, J. Chem. Phys. {\bf 56}, 5527 (1972).
\bibitem{Barbero}
G. Barbero, N. V. Madhusudana and C. Oldano, J. Phys. (France) {\bf 50}, 2263 (1989).
\bibitem{Zumer}
G. Ska$\breve{c}$ej, A. L. Alexe-Ionescu, G. Barbero and S. $\breve{Z}$umer, 
Phys. Rev. E {\bf 57}, 1780 (1998).
\bibitem{Poniewierski}
A. Poniewierski and A. Samborski, Liq. Cryst. {\bf 27}, 1285 (2000)
\bibitem{Robledo}
J. Quintana and A. Robledo, Physica A {\bf 248}, 28 (1998).
\bibitem{Sullivan2} 
A.K. Sen and D.E. Sullivan, 1987, Phys. Rev. A {\bf 35}, 1391. 
\bibitem{Evans}
R. van Roij, M. Dijkstra and R. Evans, Europhysics Letters {\bf 49}, 350 
(2000). 
\bibitem{Teixeira}
P. I. C Teixeira, Phys. Rev E {\bf 55}, 2876 (1997).
\bibitem{Inma1}
I. Rodr\'{\i}guez-Ponce, J.M Romero-Enrique, E. Velasco, L. Mederos and
L. F. Rull, Phys. Rev. Lett. {\bf 82}, 2697 (1999). 
\bibitem{Inma2}
I. Rodr\'{\i}guez-Ponce, J. M. Romero-Enrique, E. Velasco, L. Mederos and
L. F. Rull, J. Phys.: Condens. Matter {\bf 12}, A367 (2000).
\bibitem{Margarida}
M. M. Telo da Gama, P. Tarazona, M. P. Allen and R. Evans, Molec. Phys.
{\bf 71}, 801 (1990).
\bibitem{anchoring} F. N. Braun, T. J. Sluckin, E. Velasco and
L. Mederos, Phys. Rev. E {\bf 53}, 706 (1996). An alternative
numerical approach is that used in E. Mart\'{\i}n del R\'{\i}o, M. M. Telo da
Gama, E. de Miguel and L. Rull, Phys. Rev. E {\bf 52}, 5028 (1995). 
\bibitem{minimo} Y. Mart\'{\i}nez, E. Velasco, A. M. Somoza, L. Mederos
and T. J. Sluckin, J. Chem. Phys. {\bf 108}, 2583 (1998).          
\bibitem{Cleaver}
G. D. Wall and D. J. Cleaver, Phys. Rev. E {\bf 56}, 4306 (1997).
\bibitem{Oseen} C. W. Oseen, Trans. Faraday Soc. {\bf 29}, 833 (1933).
\bibitem{Frank} F. C. Frank, Discuss. Faraday Soc. {\bf 25}, 19 (1958).
\bibitem{Yokoyama} H. Yokoyama, Phys. Rev. E {\bf 55}, 2938 (1997).
\bibitem{Faetti} M. Faetti and S. Faetti, Phys. Rev. E {\bf 57}, 6741 (1998).
\bibitem{TeixeiraJCP}
P. I. C. Teixeira, J. Chem. Phys. {\bf 97}, 1498 (1992).
\endgroup
\end{thebibliography}
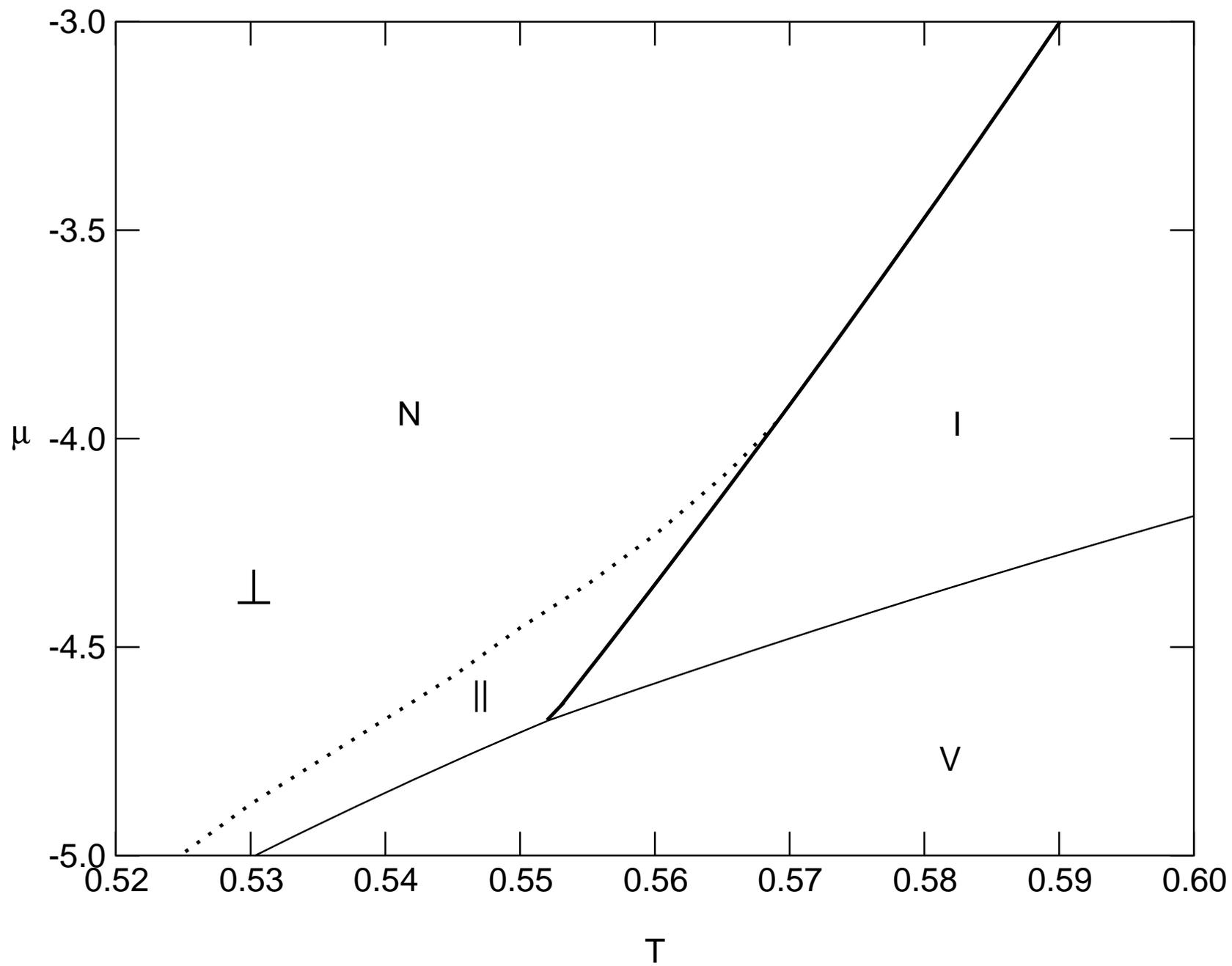
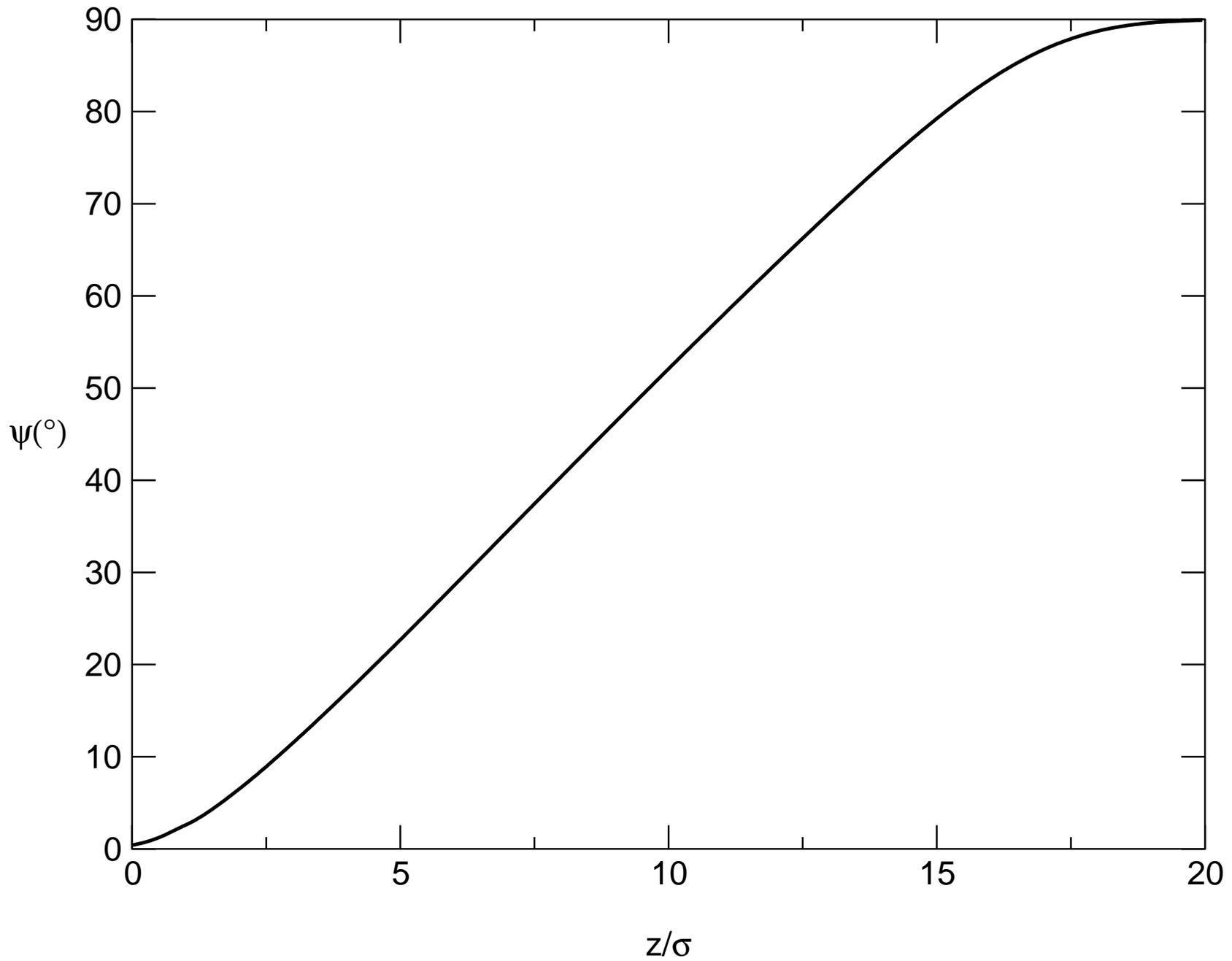
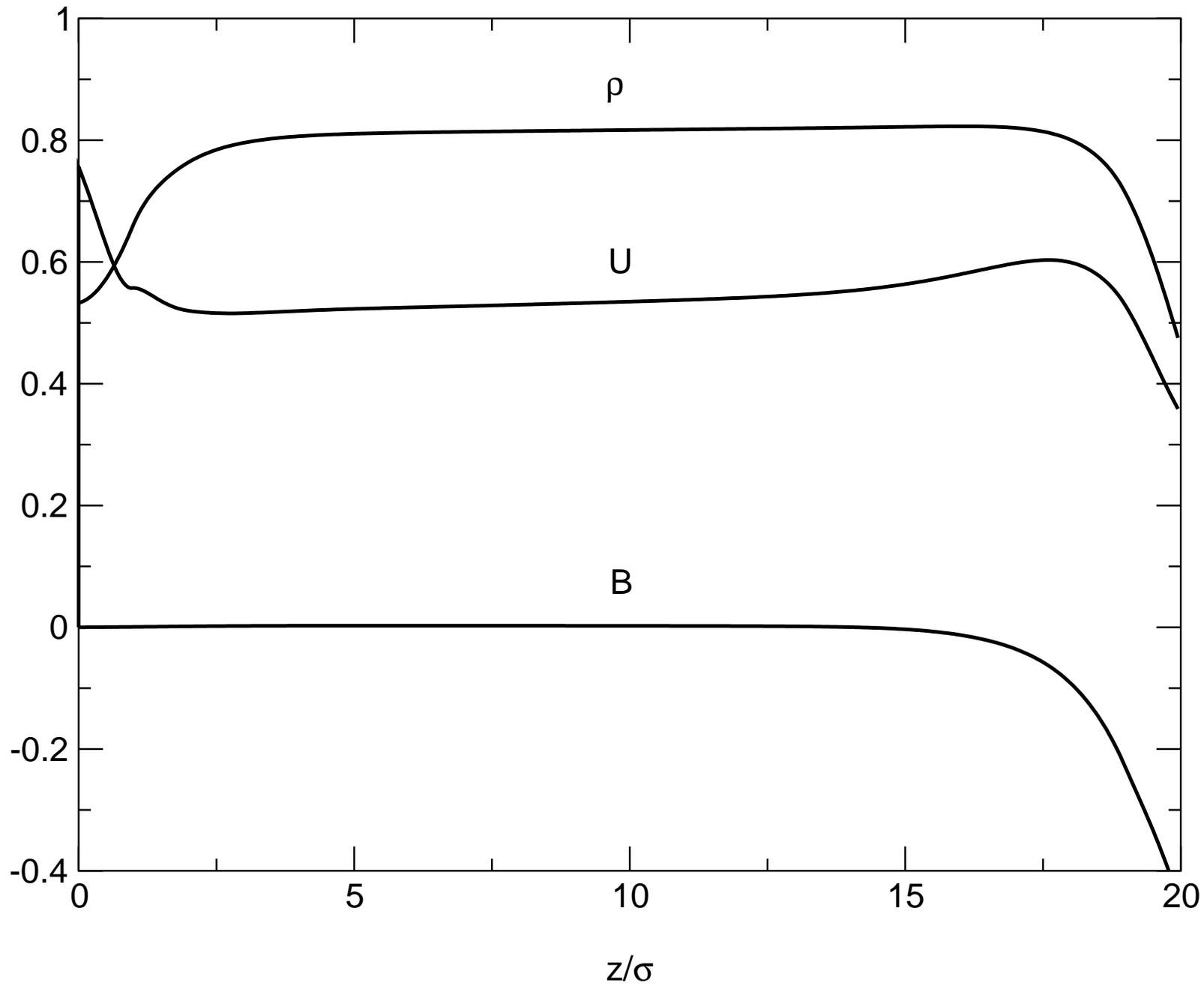
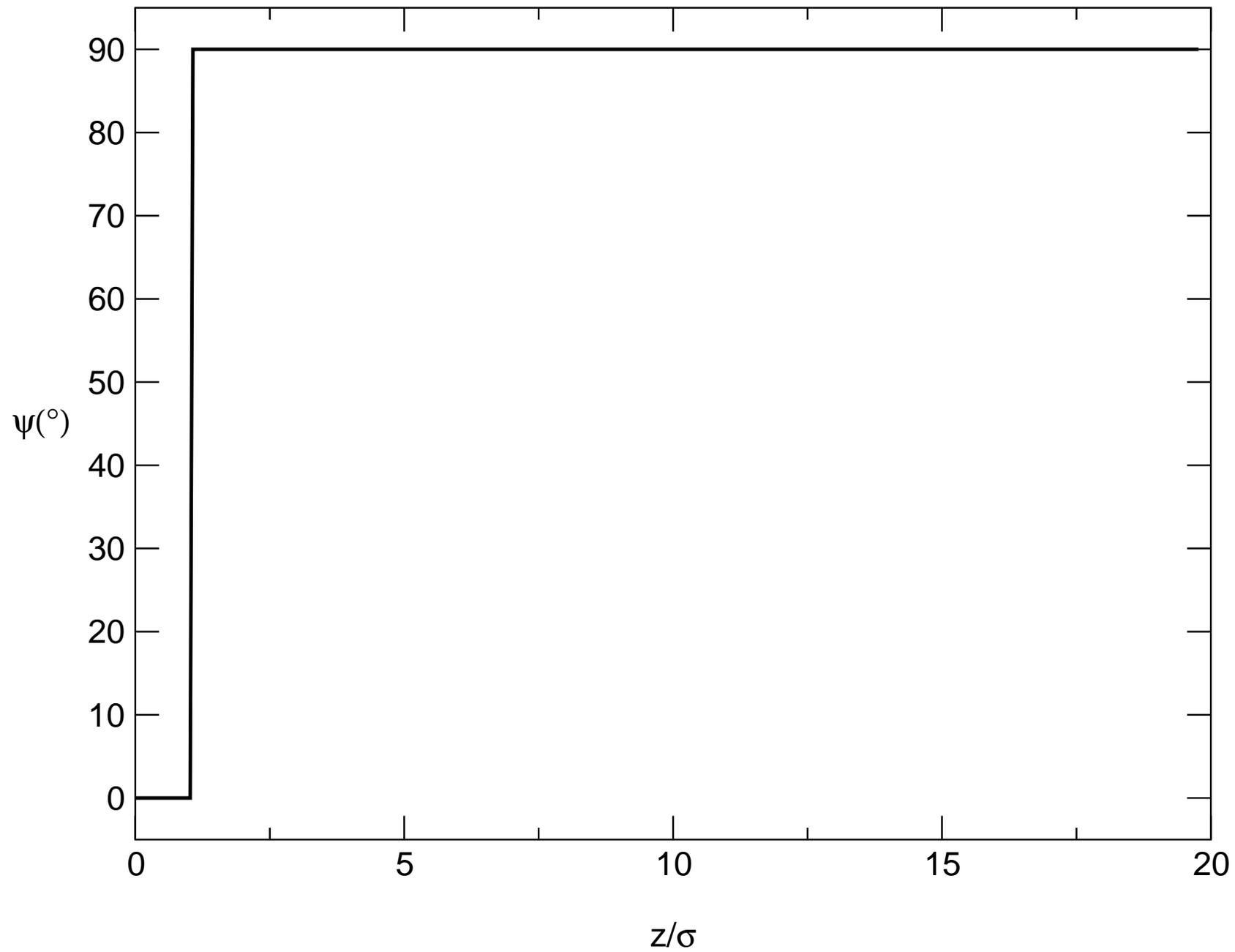
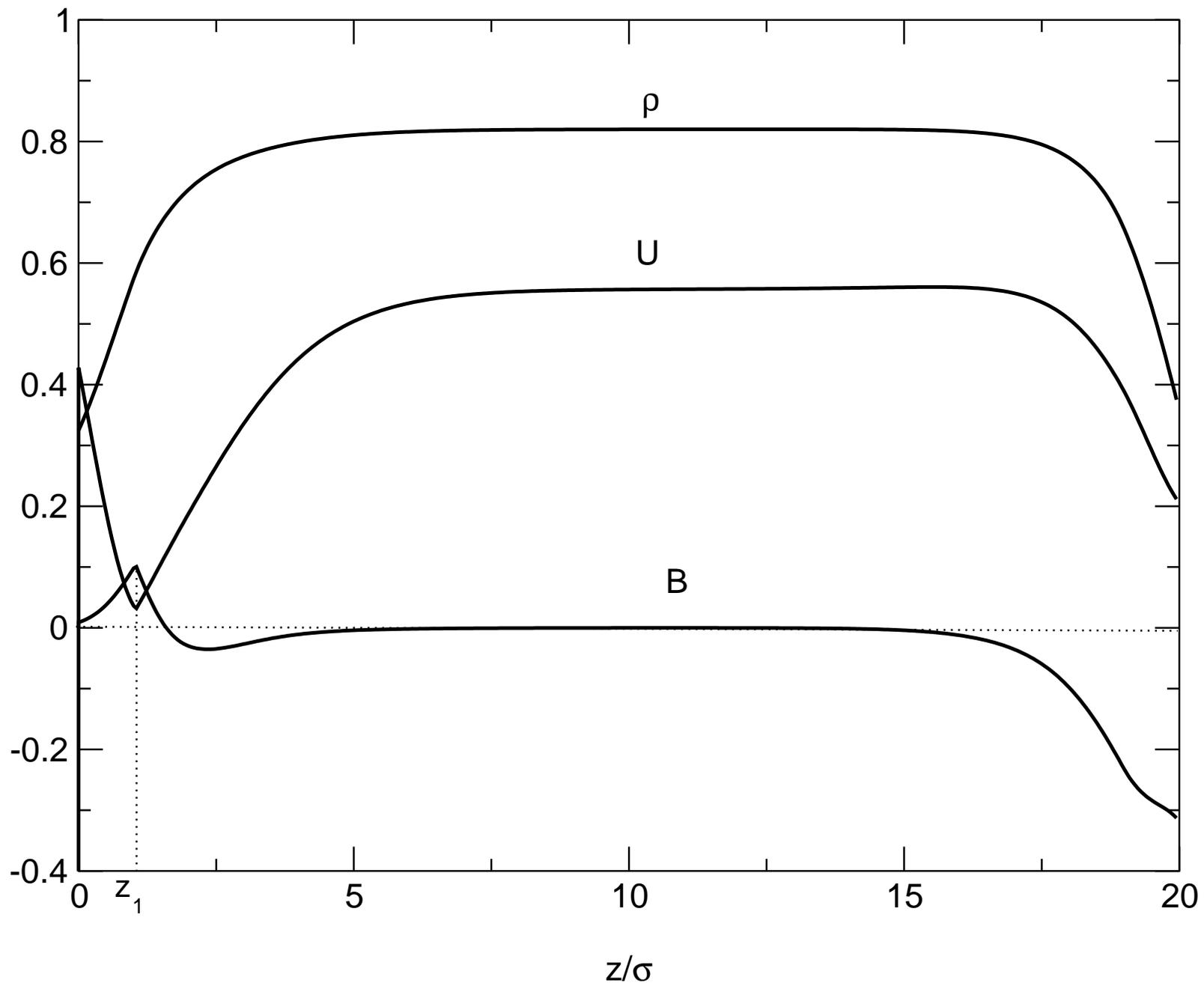
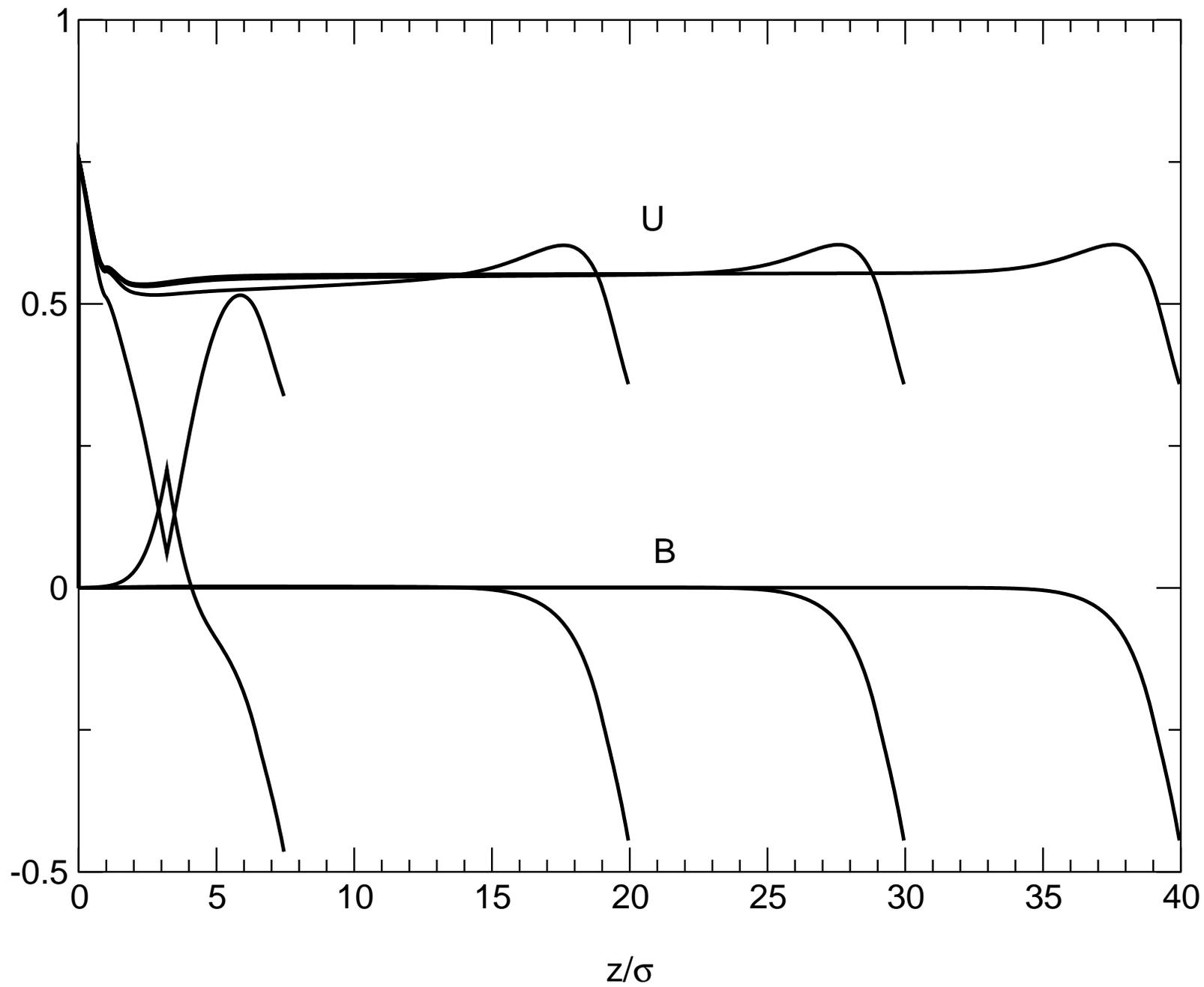
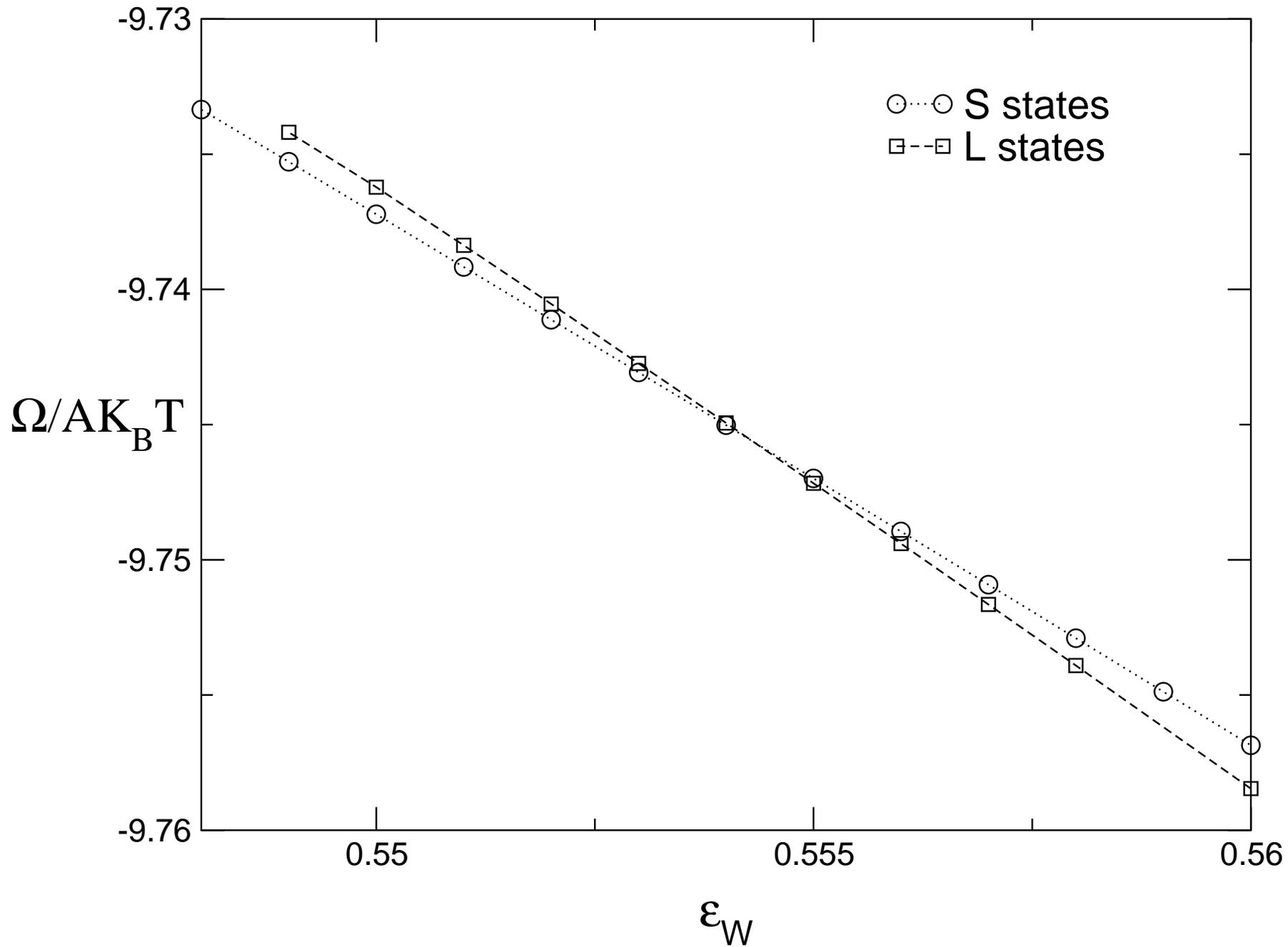
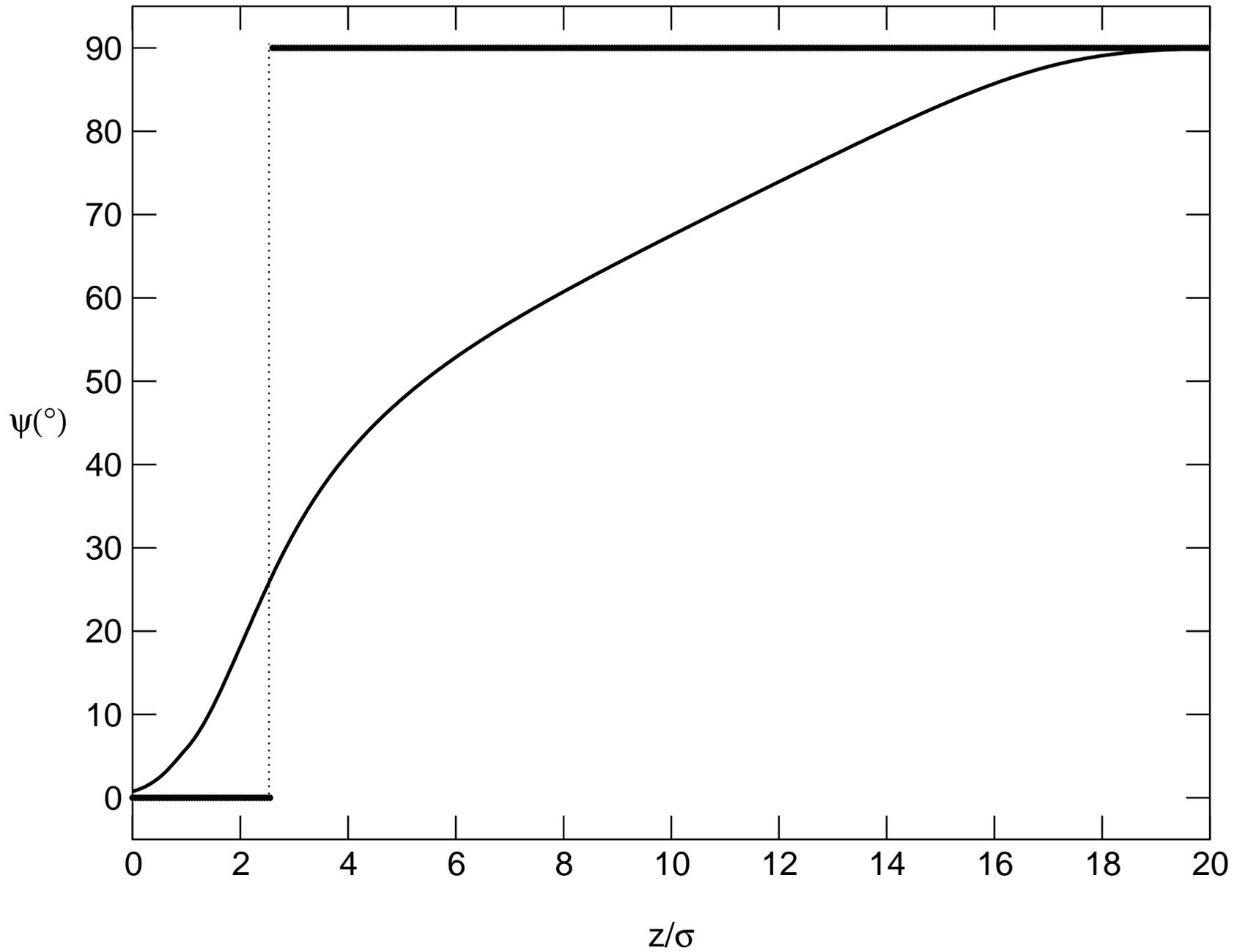
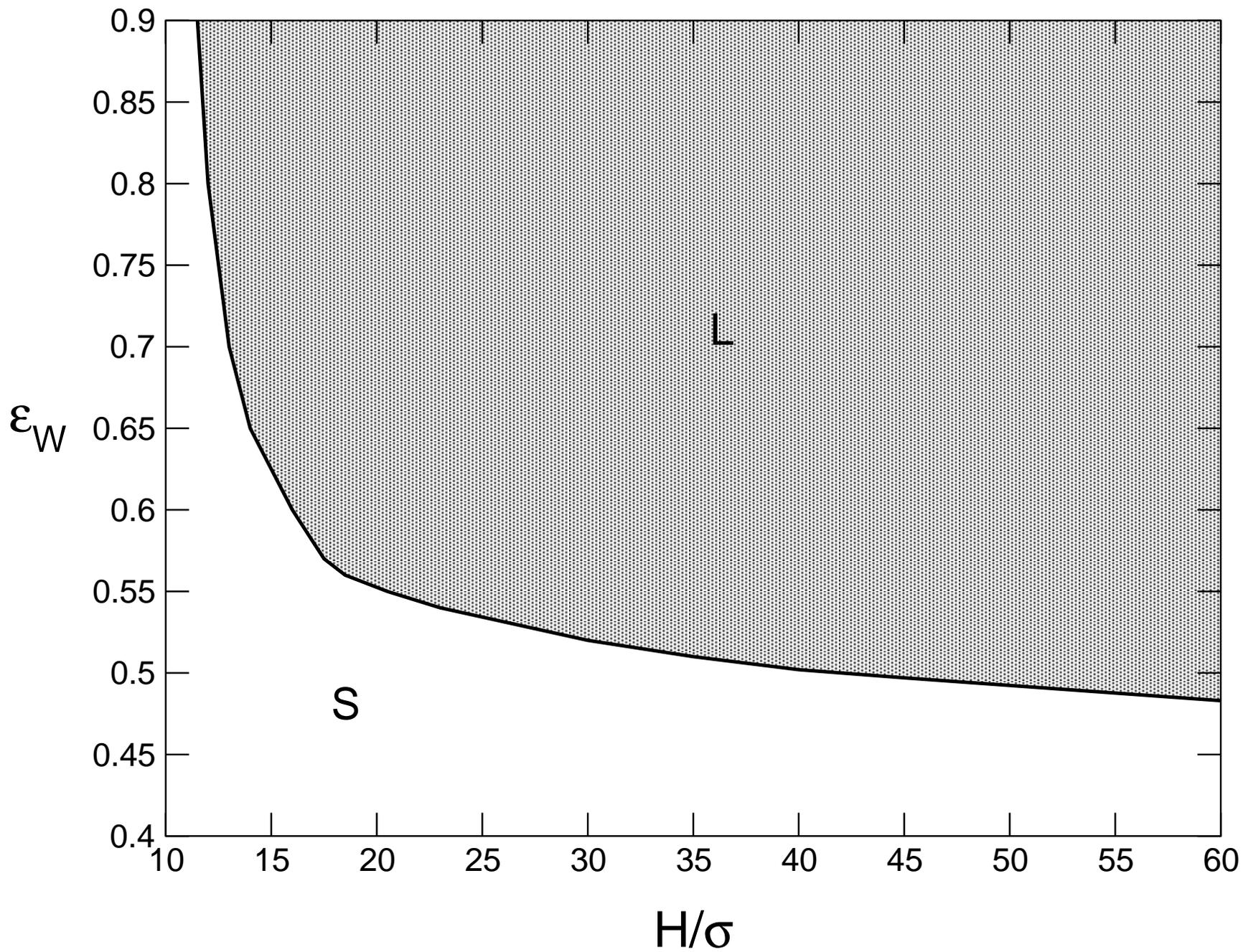
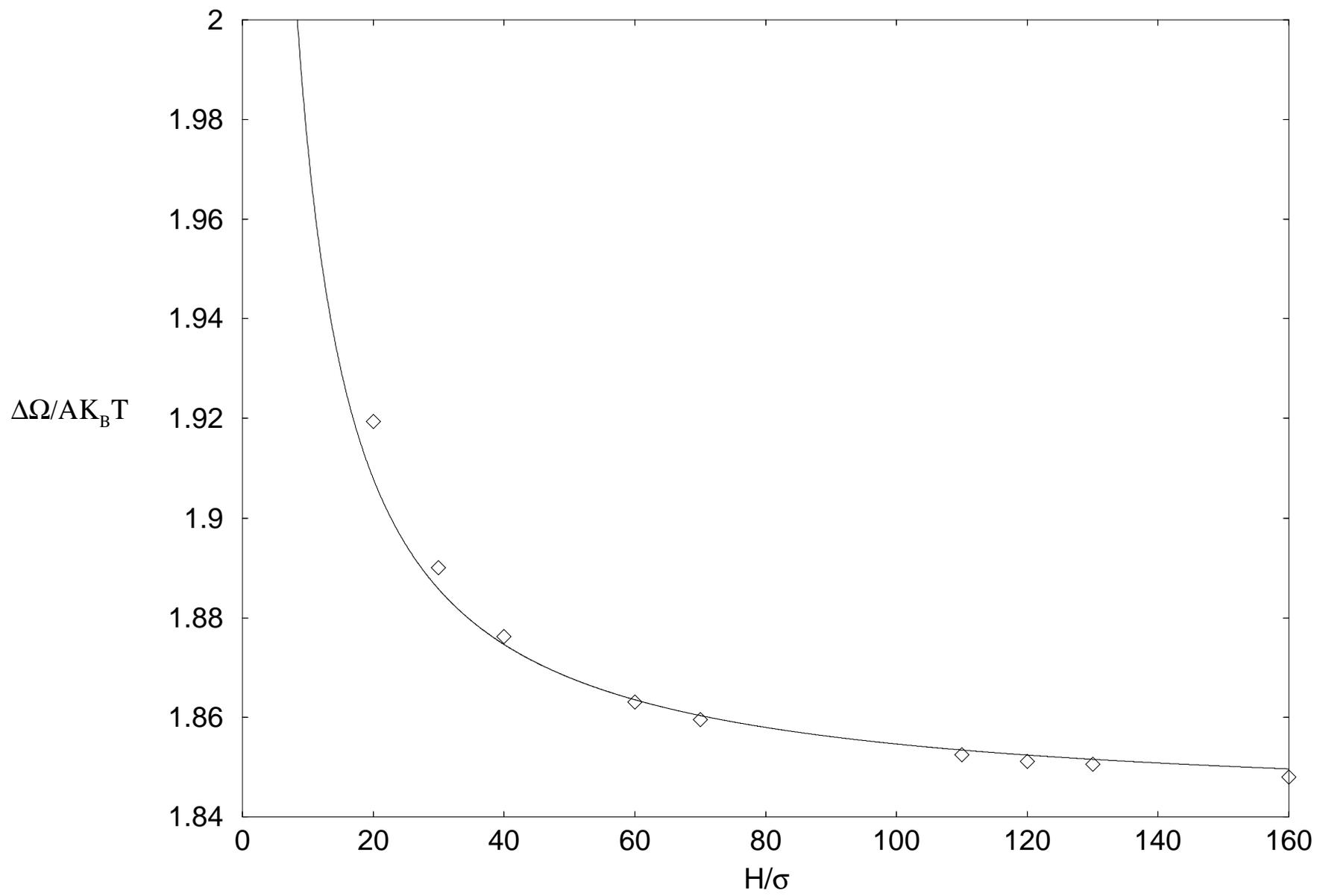
\end{document}